# Diffuse neutron reflectivity and AFM study of interface morphology of an electro-deposited Ni/Cu film

Surendra Singh[1], Saibal Basu[1] and S. K. Ghosh[2]

[1]Solid State Physics Division, Bhabha Atomic Research Center, Mumbai-85, India
[2]Material Processing Division, Bhabha Atomic Research Center, Mumbai-85, India

## Abstract

We present a detailed study of the interface morphology of an electro-deposited (ED) Ni/Cu bilayer film by using off-specular (diffuse) neutron reflectivity technique and Atomic Force Microscopy (AFM). The Ni/Cu bilayer has been electro-deposited on seed layers of Ti/Cu. These two seed layers were deposited by magnetron sputtering. The depth profile of density in the sample has been obtained from specular neutron reflectivity data. AFM image of the air-film interface shows that the surface is covered by globular islands of different sizes. The AFM height distribution of the surface clearly shows two peaks [Fig. 3] and the relief structure (islands) on the surface in the film can be treated as a quasi-two-level random rough surface structure. We have demonstrated that the detailed morphology of air-film interfaces, the quasi-two level surface structure as well as morphology of the buried interfaces can be obtained from off-specular neutron reflectivity data. We have shown from AFM and off-specular neutron reflectivity data that the morphologies of electro-deposited surface is distinctly different from that of sputter-deposited interface in this sample. To the best of our knowledge this is the first attempt to microscopically quantify the differences in morphologies of metallic interfaces deposited by two different techniques viz. electro-deposition and sputtering.

Key words: Neutron reflectivity, Specular, Off-specular, Interface roughness, AFM, Morphology.

## 1 INTRODUCTION

X-ray and neutron reflectometry, as powerful non-destructive techniques, have been used to investigate interfaces with respect to roughness, compositional and density gradients at free surfaces and at the interfaces of multilayers [1-6]. From specular reflectivity, one can obtain the laterally averaged density profile of a sample in the direction normal to its surface. In other words one obtains one dimensional density gradient from specular reflectivity. Off-specular x-ray and neutron reflectivity are increasingly being used to study the in-plane structure or morphology of the surfaces [6-10]. There are several imaging techniques, e.g., Scanning Tunneling Microscopy (STM), Atom Force Microscopy (AFM), etc., which allow one to view the film surface by mapping the height of the surface at the air-film interface. Off-specular x-ray and neutron scattering provide global statistical information about the spatial correlations parallel to the surface of a material over the coherent length of the beam. Neutrons being a neutral particle and due to their deep penetrability are able to provide information of the buried interfaces in a sample. The application of off-specular x-ray and neutron reflectivity have become widespread, including study of capillary-wave fluctuations at a liquid surface [11], partially correlated interface roughness in multilayers [8, 9, 10, 12], laterally modulated structure of periodic gratings [13, 14], morphology of diblock copolymers [15-18] and other systems [19-21]. Detailed understanding of in-plane morphology and roughness of interfaces of a layer system is required to understand the influence of these factors on magnetic, electric and optical properties of layered systems [22-24]. Also morphological study of thin films is of great importance for a better understanding of the physical mechanism of thin film growth and for preparing thin films with tailored morphology in order to acquire unique physical and chemical properties [25, 26]. For example, the morphology of transition metal film is of critical importance for the growth of carbon nanotubes [25-27] and it has played major role to control the density, diameter, and length of the aligned carbon nanotubes. Specular and off-specular x-ray scattering has been successfully applied to study the morphology of films, which are covered by the relief structures (i.e., holes or islands) on their surfaces [16-18].

In a previous study, using off-specular neutron reflectivity we compared the change in morphology of the surface caused by corrosion with respect to a buried interface in a Ni film grown on glass substrate using thermal evaporation technique [28]. In the present work we have attempted to unravel the detailed morphology of the interfaces in a Ni/Cu sample, grown by electro-deposition (ED) method, using specular neutron reflectivity (NR), off-specular (diffuse) neutron reflectometry (DNR) and AFM. ED is a widely used commercial technique for deposition of protective coatings over metallic surfaces. In comparison to other methods, ED is simple and less expensive. Over last few years ED has been shown to be a valuable technique for obtaining metallic bilayers and multilayers [29, 30]. Earlier, we found from AFM study on films deposited by different techniques [31] that the exposed surface of an ED film shows very distinctive morphology vis-à-vis vacuum deposited films. The present sample is a Ni/Cu film grown by ED. The air-film interface has been probed by AFM. The surface is covered with three dimensional islands [Fig. 2(A)]. Presence of such islands with varying heights on the surface of the sample has been treated as a quasi-two-level random rough surface in off-specular neutron reflectivity analysis in the present work. We have analyzed the scaling behavior of height-height correlation function on the surface and have obtained roughness, roughness exponent and the correlation length at the air-film interface of the sample from the AFM measurements as well as from off-specular neutron reflectivity data. From detailed analysis of the off-specular neutron reflectivity data we have been able to show that the morphology of the buried interfaces, grown by sputtering and the electro-deposited interface, are different. Also we have shown that the quasi-two level height distribution of the exposed surface has given rise to an intensity oscillation in the DNR data.

## 2 SAMPLE PREPARATION AND EXPERIMENTAL

The film was grown on Si (111) substrate at room temperature. The Ni/Cu bilayer sample was electrodeposited on a 100 Å thick Cu seed layer. The seed layer in turn was deposited on a buffer layer of Ti of thickness 250 Å. Both seed and buffer layers were grown by magnetron sputtering methods. Si wafer was etched in 5% HF solution prior to deposition. Ni/Cu bilayer film was pulse-electrodeposited from a single solution electrolyte. The electrolyte used was: 0.7 M $NiSO_4.7H_2O$, 0.02 M $CuSO_4.5H_2O$ and 0.2 M Sodium citrate. All the chemicals used were of analytical grade. Solutions were prepared using de-ionized water from Millipore Milli-Q system. The pH of the electrolyte was kept constant at 6.0 throughout the deposition. Deposition was carried out by galvanostatic method using a fabricated three-wave pulse generator from a quiescent solution at 25 ± 1 $^0$C. The pulse generator basically generates square-wave pulses. Copper was deposited at current density (C.D.) of 0.2 mA.cm$^{-2}$ while the (C.D.) for nickel was 20 mA.cm$^{-2}$ to minimize copper content at the deposited Ni layer.

Specular and off-specular neutron reflectometry measurements at room temperature were performed at the polarized neutron reflectometer at Dhruva, Mumbai, India [32]. The schematic of the polarized neutron reflectometer at Dhruva is shown in Fig 1. This instrument has been designed for vertical sample geometry with a horizontal ${}_2He^3$-based linear position sensitive detector (PSD), located normal to the incident beam to capture the reflected intensity. The specular reflectivity pattern was collected as a function of momentum transfer $q_z$ in Å$^{-1}$, normal to the plane of the thin film, in a step-scan mode by rotating the vertical sample. The off-specular data can be collected in the horizontal plane along the position sensitive detector. This unique arrangement allows us to collect off-specular data simultaneously over all the channels of the PSD, at any $q_z$ value. Length of the PSD restricts the range of momentum transfer in DNR ($q_x$ in this case). Inset in Fig. 1 shows the wave vectors of the incident and scattered neutron are $K_I$ and $K_F$, with the incidence and exit angle $\theta_I$ and $\theta_F$, respectively. The momentum transfer, $Q =.K_F - K_I = (q_x, 0, q_z)^T$ is given by $q_x = K_I(\cos\theta_I - \cos\theta_F)$ and $q_z = K_I(\sin\theta_I + \sin\theta_F)$. Therefore a reflectivity ($\theta_I = \theta_F$) corresponds to a $q_z$ scan with $q_x$=0. The horizontal PSD collected the diffuse scattering data along its length (x-direction), integrated over vertical direction (y-direction), centered on a specular peak. The scattering geometry in our off-specular mode is equivalent to the detector scan geometry in reference [9]. The surface morphology of the sample was investigated using a NT-NDT's Solver P-47 H multimode AFM instrument. The images were taken in a non-contact mode with a $Si_3N_4$ tip. The cantilever has a spring constant of 0.6 N/m. AFM topographic images were recorded over scan areas from 1 µm × 1 µm to 30 µm × 30 µm, each with a resolution of 512×512 data points. For analyzing the specular neutron reflectometry data we have used a Genetic Algorithm (GA) based $\chi^2$ minimization program [33], which uses a matrix method [5] for generating the reflectivity pattern for a given set of physical parameters of the system.

## 3 RESULTS AND DISCUSSION

### (A) AFM

Fig. 2(A) shows the AFM image of the air-film interface of ED sample over a scan area of 5.0 µm × 5.0 µm. It shows that the film has islands of various grain sizes. For comparison, we show a typical sputter-deposited Ni film surface in Fig. 1(B). These AFM images of the two interfaces clearly demonstrate the distinct morphologies resulting due to difference in mechanism of vacuum deposition and electro-deposition. While the sputter-deposited interface shows more corrugation, but no clear grain boundaries, the ED film is covered with large globular grains of various sizes. We have shown by neutron reflectometry and AFM that the granular growth of the ED film has resulted in a very rough surface compared to vacuum-deposited interfaces. Fig. 3 (A) and 3(B) show the height distribution (histogram) of the AFM images shown in Fig. 2(A) and 2(B) respectively. The height distribution for the sputtered film [Fig 3 (B)] shows a Gaussian distribution with a single peak. The height distribution for the ED film [Fig. 3(A) ] is quite different. The histogram has been fitted with two Gaussians centered on heights of 310 Å and 630 Å. This suggests that the morphology can be described by a quasi-two level surface with two mean heights and distributions. For an exact two-level surface, one would have two δ-function peaks in the height distribution. In the present case the fraction of the upper level distribution around a height of 630 Å is 49% of the total area obtained from the Gaussian fits to the AFM height distribution. The average fractions of the upper level for AFM images taken at different regions of the film are close to this value, which indicates that the statistical distribution of heights is uniform over the entire sample. We have shown later that this has resulted in distinct oscillation in off-specular neutron reflectivity intensity. We obtained the densities, thickness of the layers and interface roughness of this film from specular neutron reflectivity data. Subsequently DNR measurements have been performed to study the morphology of the air-film interface as well as the other interfaces in the sample.

**(B) Specular neutron reflectometry**

In specular neutron reflectivity profile one observes intensity of neutrons reflected from the surface of the thin film as a function of or momentum transfer Q. Theoretical aspects of specular NR is available in references [5, 34-36]. Unpolarized NR pattern of the present sample is shown in Fig. 4. The open circles and the continuous line are the experimental data and best fit to measured data respectively. The parameters that one extracts from specular neutron reflectivity are: layer thickness, density and interface roughness. These parameters extracted from the unpolarized NR on the present sample are listed in table 1. The thickness of Cu and Ni layers grown by ED are 350 ± 15Å and 273 ± 16 Å respectively, obtained from the best fit to NR. The densities of the Ni and Cu layers, grown by ED, are 75% and 95% of their bulk values respectively. The interface roughness of the electro-deposited layers at Cu/Ni and Ni-air interfaces are 32 ± 4 Å and 36 ± 3 Å respectively. These roughness values are much larger than the roughness we obtained for the sputter-deposited layers (~9 Å) in the same film. The interface roughness parameter extracted from NR is an effective roughness, which comprises the effect of true roughness as well as that of alloying and inter-diffusion [31, 32] and doesn't tell about the true roughness and morphology of the interface. The low density of Ni layer suggests that the film has grown with considerable amount of porosity. This is due to the mechanism of electro-deposition of the Ni layer. Since, the Ni layer is deposited at a large current-density (20 $mA.cm^{-2}$) compared to Cu layer (0.2 $mA.cm^{-2}$), there is rapid hydrogen evolution at the cathode, which causes porosity in the Ni layer. The AFM image of the surface of Ni film also shows that the film surface is covered with grains of different sizes with porous regions in-between.

### (C) Off-specular neutron reflectometry: interface morphology

Detailed information about the morphology of the interfaces, i.e., the in-plane height-height autocorrelation functions have been obtained from diffuse scattering data. The diffuse scattering cross section of a single surface may be obtained using the Distorted-Wave Born Approximation (DWBA) [6, 37]. These results can be extended to layered systems including vertical correlation between interfaces [8, 38]. In the present sample the Cu and Ni layers are about 300 Å thick and vertical correlation has been neglected between the interfaces. The diffuse scattering cross-section from a multilayer system under DWBA is given by [8-10, 38]

$$\left(\frac{d\sigma}{d\Omega}\right)_{diff} = \frac{I_L K^4}{16\pi^2} \sum_{j,k=1}^{N} (n_j^2 - n_{j+1}^2)(n_k^2 - n_{k+1}^2)^* \sum_{m,n=0}^{3} \tilde{G}_j^m \tilde{G}_k^{n*} \exp\{-\frac{1}{2}[(q_{z,j}^m \sigma_j)^2 + (q_{z,k}^{n*}\sigma_k)^2]\} S_{jk}^{mn}(q_x, q_{z,j}^m, q_{z,k}^n) \quad (1)$$

Where the structure factor

$$S_{jk}^{mn}(q_x, q_{z,j}^m, q_{z,k}^{n*}) = \frac{1}{q_{z,j}^m q_{z,k}^{n*}} \int_0^\infty [\exp\{q_{z,j}^m q_{z,k}^{n*} C_{jk}(x)\} - 1]\cos(q_x x)dx \quad (2)$$

The illuminated area of the sample is denoted by $I_L$. $K$ *and* $n_j$ are the wave vector of incident neutron in air and the refractive index of the material beneath the $j^{th}$ interface. The dynamical effects [8, 9, 38] are taken into account by the factors $\tilde{G}_j^m = G_j^m \exp(-iq_{z,j}^m z_j)$, where the expression for $G_j^m$ represents the transmitted (*T*) and reflected (*R*) neutron beam at the j[th] interface: $G_j^0 = T_j^s T_j^d$; $G_j^{1} = T_j^s R_j^d$; $G_j^2 = R_j^s T_j^d$; $G_j^3 = R_j^s R_j^d$ with s and d denoting source and detector directions and $q_{z,j}^m$ are the normal components of momentum transfer vector at the j[th] interface with $q_{z,j}^0 = k_{z,j}^s + k_{z,j}^d$; $q_{z,j}^1 = k_{z,j}^s - k_{z,j}^d$; $q_{z,j}^2 = -q_{z,j}^1$; $q_{z,j}^3 = -q_{z,j}^0$; $N$ is the number of interfaces, σ the root-mean-square (rms) roughness. Our incident beam is a line beam of height about 40 mm (y-direction). The in-plane morphology of an interface is obtained from the height-height autocorrelation functions $C_j(x)$ = $C_{jj}(x)$ for an one-dimensional line beam, integrated along y axis and the cross-correlation functions $C_{jk}(x)$ gives the correlation between interfaces *j* and *k*. It is known that for many isotropic solid surfaces, $C_j(x)$ can be represented by the correlation function of a self-affine fractal surface: $C_j(x) = \sigma_j^2 e^{-(\frac{x}{\xi_j})^{2h_j}}$ with a lateral correlation length $\xi_j$ and the Hurst parameter $h_j$ [39]. The quantity $\xi_j$ describes the correlation length beyond which the height fluctuations are uncorrelated. The Hurst parameter $h_j$ is restricted to the region $0<h_j<1$ and defines the fractal box dimension $D=3-h_j$ of the interface. Small values of $h_j$ describe jagged surfaces while $h\approx1$ leads to interfaces with smooth hills and valleys and nearly two-dimensional [6, 9, 10], though both the surfaces may have same roughness parameter σ.

Fig. 5 shows the DNR pattern of the sample at $q_z$ of 0.028 $Å^{-1}$. Open circles are the measured DNR data. The discontinuous line in Fig. 5 shows the best fit obtained using Eq. (1). The parameters for different interfaces, for the above-mentioned self-affine fractal morphology are given in Table 2. The parameters obtained for various interfaces allow us to compare differences in morphology of the interfaces grown by sputtering and by electro-deposition. The parameters *h* for the sputter-deposited layers i.e., the Si/Ti interface and Ti/Cu interface above it have values of 0.48 and 0.59. In electro-deposited Cu/Ni interface and Ni/air interface this parameter jumps to 0.85 and 0.93 respectively. The same trend is seen in the fitted correlation lengths, ξ for the sputtered interfaces with respect to the electro-deposited interfaces. For the sputter-deposited Si/Ti interface and Ti/Cu interfaces ξ has values 298 Å and 398 Å respectively. The correlation length increases to 900 Å and 1298 Å in the electrodeposited Cu/Ni and Ni/air interfaces respectively. These parameter values points to the basic difference in the process of thin film growth by sputtering and by electro-deposition. AFM images of the electro-deposited film in Fig. 2(A) and sputter-deposited film in Fig. 2(B) highlights this point. While the sputter-deposited interfaces have much smaller average roughness (~9 Å) in comparison with the electro-deposited film (36 Å), the electro-deposited interfaces are much smoother compared to sputter-deposited interfaces at short length scales. This "*jaggedness*" is apparent in the AFM image in Fig. 2(B) and results in shorter correlation length in sputter-deposited layers compared to electro-deposited layers. The average roughness parameter obtained from specular reflectivity indicates peak to trough distance at the interface and it is large for the electro-deposited layers due to the granular growth. This is indicated by a large value of roughness parameter (36 Å) at the Ni/air interface. Though the electro-deposited film is covered by grains of various sizes, the surface is smoother compared to sputter-deposited interfaces and this is also apparent from higher value of the parameter *h*.

Moreover, the interfacial width obtained from the specular reflectivity includes the effects of both interfacial roughness and interdiffusion, which can be expressed as $\sigma_{tot}^2 = \sigma_c^2 + \sigma_d^2$ , where $\sigma_c$ and $\sigma_d$ represent the true roughness (root mean square of height-height difference) and roughness due inter-diffusion respectively. The roughness measured from the off-specular reflectivity measurements is true roughness. Therefore comparing the roughness value obtained from two measurements gives the inter-diffusion width $\sigma_d$. The roughness parameters for Cu/Ni obtained from two measurements, NR and DNR, are quite different. The roughness for Cu/Ni interface obtained from NR and DNR are 32 Å and 9 Å respectively, which suggest there is considerable amount of interdiffusion ($\sigma_d \approx 31$ Å) at this interface. The roughness for Ni/air interface obtained from NR and DNR measurements are close. The roughness and grain size distribution of thin film strongly depends on the kinetics of the nucleation and growth mechanism of the technique.

The best fit for DNR data (discontinuous line in Fig. 5) assuming layered structure without conformal roughness gives a reasonable fit over the entire $q_x$ range and also provides an insight in to the basic morphological differences of sputtered layers with respect to electro-deposited layers. While the fit is justified within experimental error bars for the present data, it completely disregards the small oscillation in the experimental data at $q_x = 0.005$Å$^{-1}$. The height distribution obtained from AFM [Fig. 3(A)] clearly shows that present sample has a quasi-two level relief structure on the surface. The work of Vignaud et. al. [18] on off-specular scattering from relief structure of diblock copolymer films and that of Zhao et. al.. [40] on the relief structure due to pitting corrosion on Al film considers the correlation between the two levels in the transverse off-specular scan in terms of correlation between domains in the plane of the sample. We have added an extra term to account for the quasi-two dimensional relief structure due to the grains on the surface in addition to DWBA diffuse scattering cross-section for uncorrelated interfaces in Eq. (1). To a first approximation the total diffuse cross-section may be written as:

$$\left(\frac{d\sigma}{d\Omega}\right)_{Dtot} = \left(\frac{d\sigma}{d\Omega}\right)_{diff} + \left(\frac{d\sigma}{d\Omega}\right)_{isd} \qquad (3)$$

Where the first term on the right hand side of Eq. (3) is given in Eq. (1) and shows the nature of the roughness and morphology of all the interfaces in the sample through height-height correlation function at the respective interface. The second term in Eq. (3) is the sole contribution from the quasi-two dimensional relief structure at the air-film interface, defined as [18, 40, 41]

$$\left(\frac{d\sigma}{d\Omega}\right)_{isd} \ \alpha \ \frac{K^4}{16\pi^2}\left|1-n^2\right|^2 \Theta(1-\Theta)\sin^2\left(\frac{\sigma q_z}{2}\right)\Gamma(q_x) \qquad (4)$$

Where Θ is the fraction of the upper level area, which has been determined from AFM image of the air-film interface, σ is the roughness of the air-film interface and the integration $\Gamma$ ($q_x$) is Fourier transform of the Bessel function that defines correlation between the two levels on the surface.

$$\Gamma(q_x) = \sum_l r_l e^{-(r_l/\xi)^{2h}} j_0(q_x r_l) \qquad (5)$$

The continuous line shown in Fig. 5 represents the best fit obtained using the Eq. (3), in which we have incorporated the relief surface contribution as described above. One can see that a good agreement exists between the calculated and the measured intensity, for whole $q_x$ range, when we consider a quasi-two level approach. The small oscillation at $q_x = 0.005$Å$^{-1}$ in the DNR data may be associated with a lateral distribution of grain size with an average size of 1260 Å ($2\pi/q_x$). However the large width of the peak along $q_x$ indicates a broad distribution of grain size around this average size.

**(D) A comparison of DNR and AFM measurements**

In the present analysis for the morphology at air-film interface, we have used AFM data from the sample over scan areas ranging from 1.0μ × 1.0μ to 30.0μ × 30.0μ with space resolution varying from ~20 Å to 600 Å respectively. The average surface topography has been obtained by calculating root-mean-square (RMS) roughness from height measurement in AFM. The roughness calculation is simplest and the most commonly used parameter for observation of surface morphology. We found the RMS roughness increases from 45 Å to 230 Å as we go to higher scan area. The RMS roughness is strongly dependent on the scan size [42, 43]. This confirms that the sample surface has fractal geometry over several length scales [44]. Surface morphology of films, may be quantified using a height-difference correlation function $g(r)$ defined as [45]

$$g(r) = \langle [z(r) - z(0)]^2 \rangle \qquad (6)$$

Where *z(r)* denotes, the height of a surface at the lateral position $r$ is determined from AFM measurements. The angular bracket denotes an ensemble average. In case of a self-affine fractal surfaces one can express *g(r)* as given below [20].

$$g(r) = 2\sigma^2[1 - e^{-(\frac{r}{\xi})^{2h}}] \qquad (7)$$

The function $g(r)$ can be calculated from the set of heights *z(r)* obtained from the AFM scan. In the present case the angular resolution for in-plane momentum transfer in the DNR, $\Delta q_x$ is approximately $10^{-4}$ Å$^{-1}$, which means a spatial range of about 11000 Å or 1.1 μ. This allows us to compare the AFM scan over 1.0μ × 1.0μ with the DNR result. The open circles in Fig. 6 are the functions *g(r)* from AFM data of 1.0μ × 1.0μ area, using Eq. (6). The solid line in this figure corresponds to the best-fit using Eq. (7). The fit gives an RMS roughness (σ) of 46.0 Å, a correlation length ($\xi$) of 1150.0 Å and the Hurst parameter (*h)* of 0.90. These values are in good agreement with the parameters for air-film interface that obtained from the DNR measurements. We also note that the AFM data was collected typically over few micrometers on the sample surface, while the DNS data gives an average picture over the entire sample surface of few centimeters. Such close match between AFM and DNS data indicates that the fractal behavior of the film surface prevails from intra-grain to inter-grain length scale on the film surface.

## 4. SUMMARY AND CONCLUSION

In summary, we have attempted a detailed characterization of the interface morphology of an electro-deposited Ni-Cu film, with a relief structure at the air-film interface, using specular neutron reflectometry, diffuse neutron reflectometry and AFM techniques. We want to emphasize that due to the difference in the inherent growth process between the techniques of electro-deposition and sputtering, the interface morphologies are quite distinct. Electrodeposited film shows a granular growth of film with a distinct quasi-two level surface morphology. The DNS data reveals the detailed morphology of buried interfaces. The interfaces grown by sputtering technique show a self-affine rough fractal surfaces with a small correlation length of $\xi \approx 400$ Å, whereas the Cu/Ni and Ni/air interface (grown by electro-deposition method) are found to be almost two dimensional with a rather large correlation length of $\xi \approx 1400$ Å, which is in the same range as the island size in lateral dimension. The presence of higher roughness for ED film is directly related to the island like growth of a multigrain structure with a high density of defects (porosity). From AFM data it is clear that the film has a relief structure at the air-film interface. We could confirm this from detailed analysis of DNR data. Also the morphological parameters obtained for the surface fractal layer from DNR and AFM are in excellent agreement.

Table 1. The structural parameters of the sample extracted from unpolarized neutron reflectometry measurements.

| Material | Thickness (Å) | Density (g/cc) | Roughness (Å) |
|---|---|---|---|
| Si (substrate) | ---- | 2.2±0.2 | 9±2 |
| Ti (buffer) | 240±12 | 3.8±0.3 | 11±2 |
| Cu | 350±15 | 8.1±0.3 | 32±4 |
| Ni | 273±16 | 6.0±0.5 | 36±3 |
| Air | ---- | ---- | ----- |

Table 2. The morphological parameters extracted from the off-specular neutron reflectivity measurements

| Material | Roughness (Å) | Correlation Length, ξ (Å) | H |
|---|---|---|---|
| Si (substrate) | 8±1 | 298±20 | 0.48±0.03 |
| Ti (buffer) | 9±1 | 398±33 | 0.59±0.04 |
| Cu | 9±2 | 900±80 | 0.85±0.05 |
| Ni | 36±3 | 1298±100 | 0.93±0.04 |
| Air | ---- | ---- | ----- |

Figure captions :

Fig. 1: Schematic of polarized neutron reflectometer at Dhruva. Inset shows the schematic of scattering geometry in the off-specular reflectivity mode. The $q_z$ arrow denotes the specular ridge along *qz*.

Fig. 2: AFM images of the Ni-air interface (A) of electrodeposition (ED) sample and a typical Ni surface grown by sputtering technique of size 5.0μm × 5.0μm recorded in non-contact mode.

Fig. 3: (A): The histogram of the height in AFM image (shown in Fig. 2 (A)) of the Ni-air interface of the ED sample. Open circles represents the height distribution of AFM image, whereas the open triangles are two Gaussian fits. The solid curve is the addition of two Gaussian fits. (B): The topographical histogram of the AFM image shown in Fig. 2 (B). The scattered data represent the height distribution from AFM image. The solid curves are the best Gaussian fits.

Fig. 4: Unpolarized neutron reflectometry profile of the sample. Open circles and continuous line represent the experimental data and the fit respectively.

Fig. 5: The DNS data for the sample at $q_z$=0.028$Å^{-1}$. Open circles are background corrected experimental data. The continuous and discontinuous lines shows the best fit with and without relief structure at surface of the film (see text).

Fig. 6: The correlation function (see text) of the AFM data of scans area 1.0μ × 1.0μ. The inset shows the AFM image of 1.0μ × 1.0μ scan size.

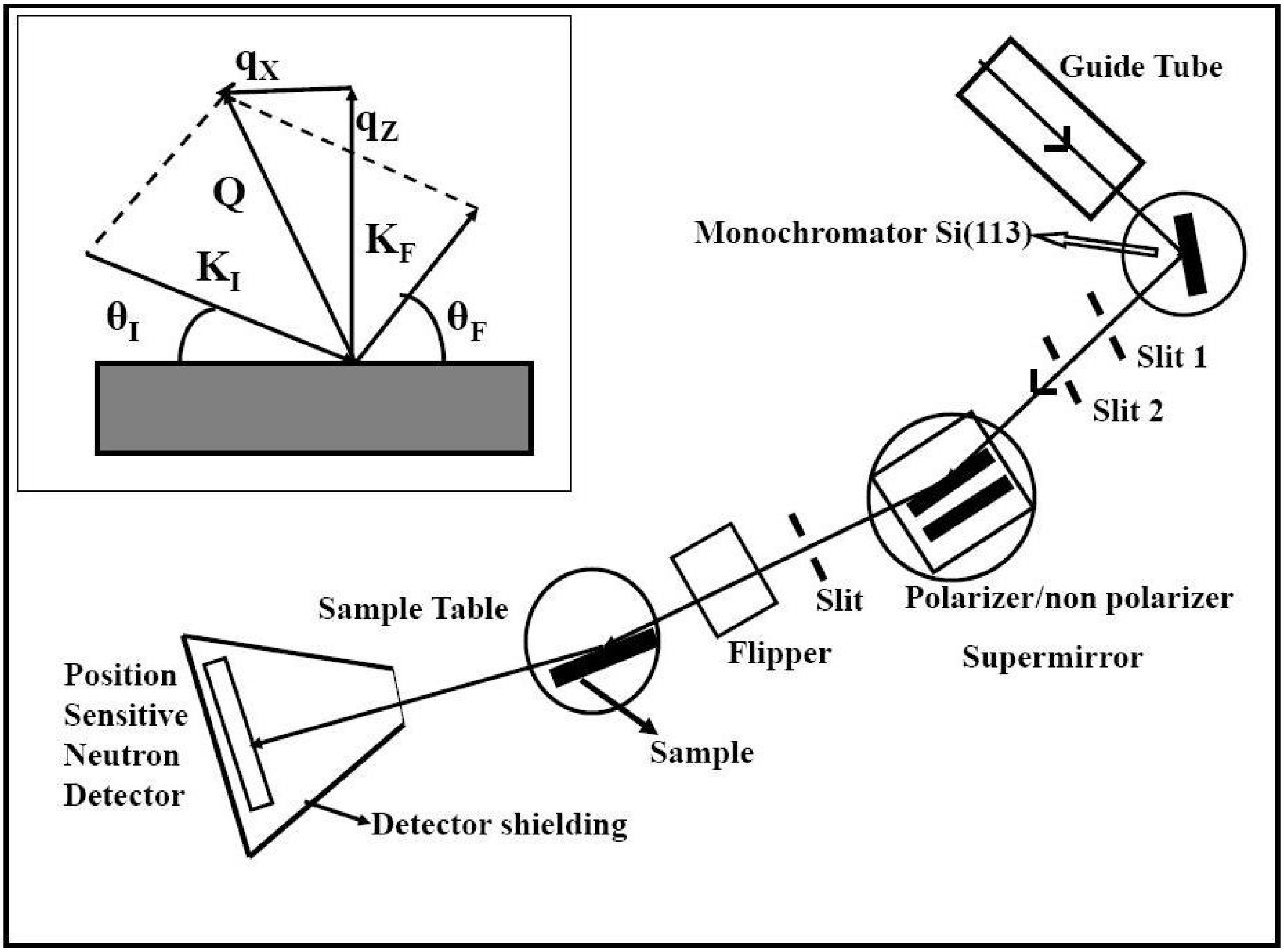
qX
qZ
Q
KF
KI
θI
θF
Guide Tube
Monochromator Si(113)
Slit 1
Slit 2
Slit
Polarizer/non polarizer
Supermirror
Flipper
Sample Table
Sample
Position
Sensitive
Neutron
Detector
Detector shielding

Fig. 1

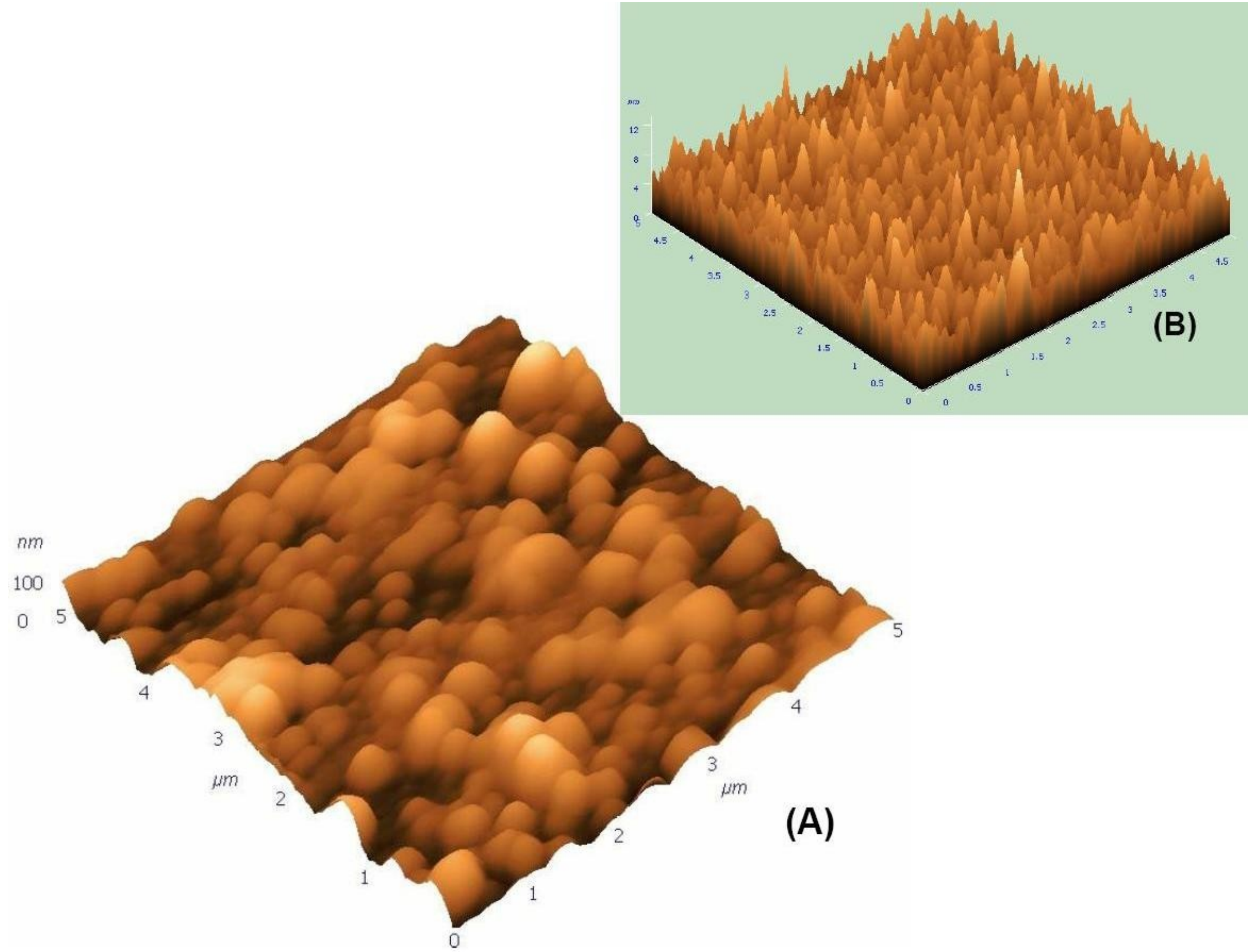

Fig. 2

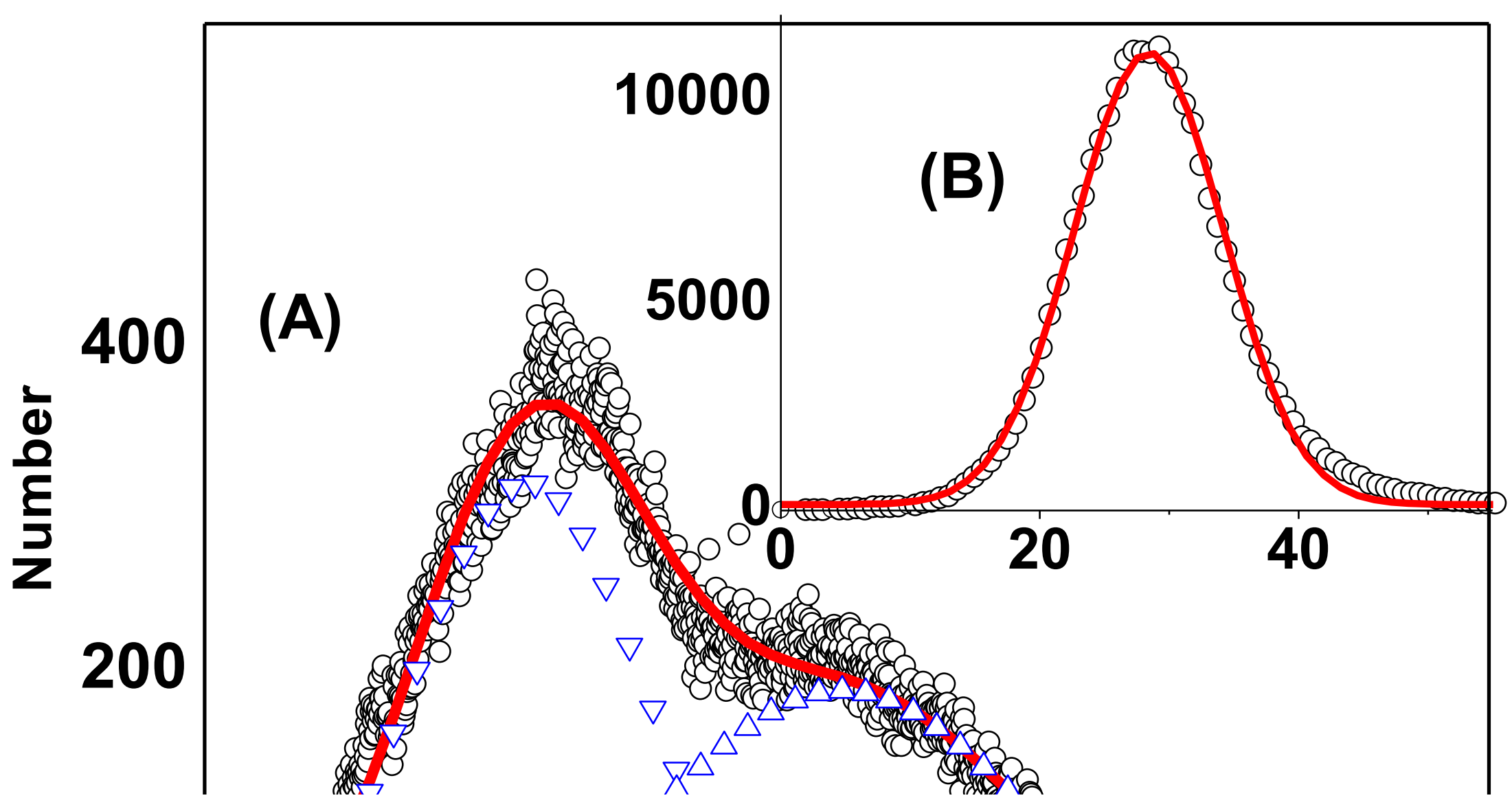

Fig 3

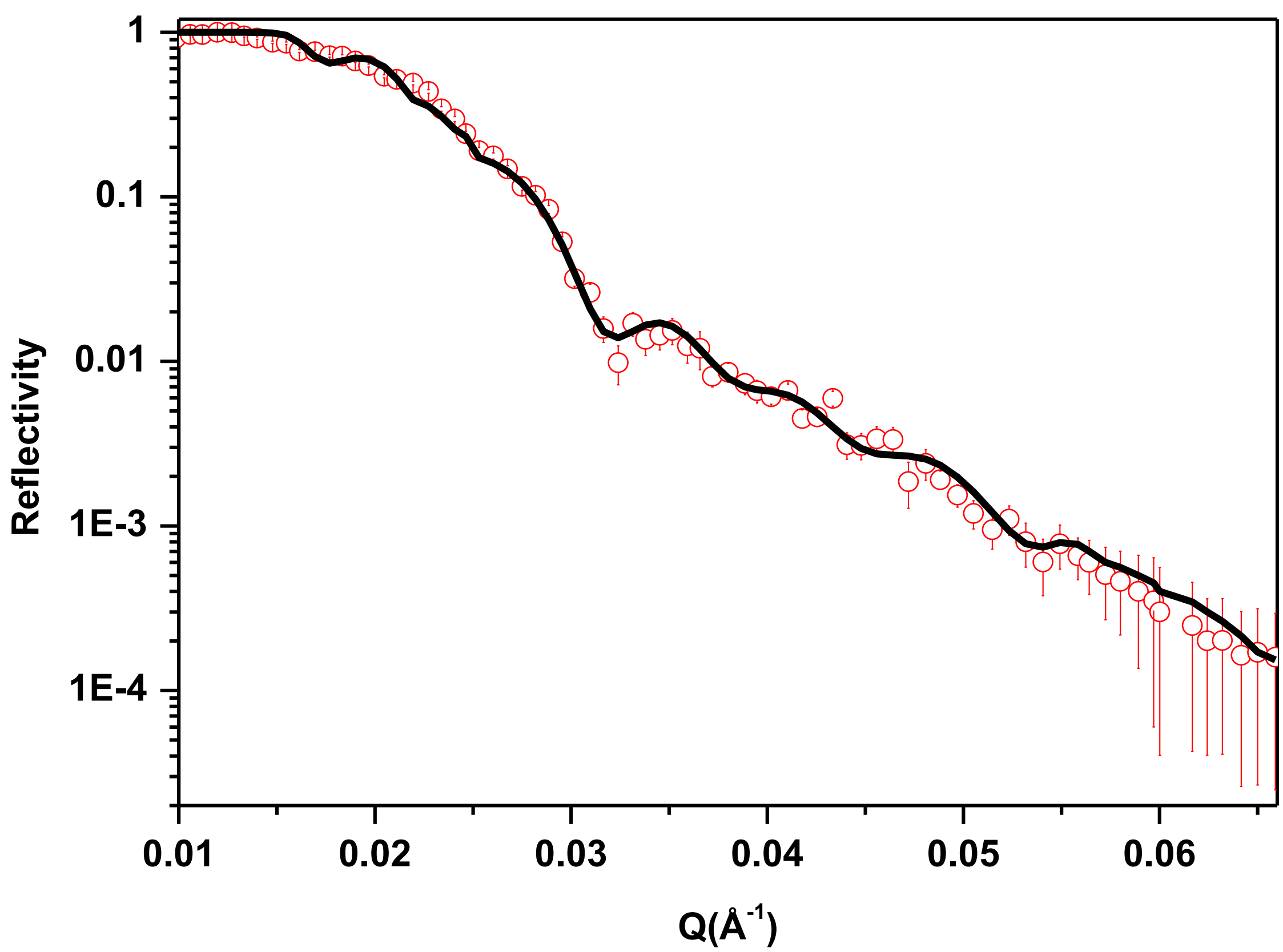

Fig. 4

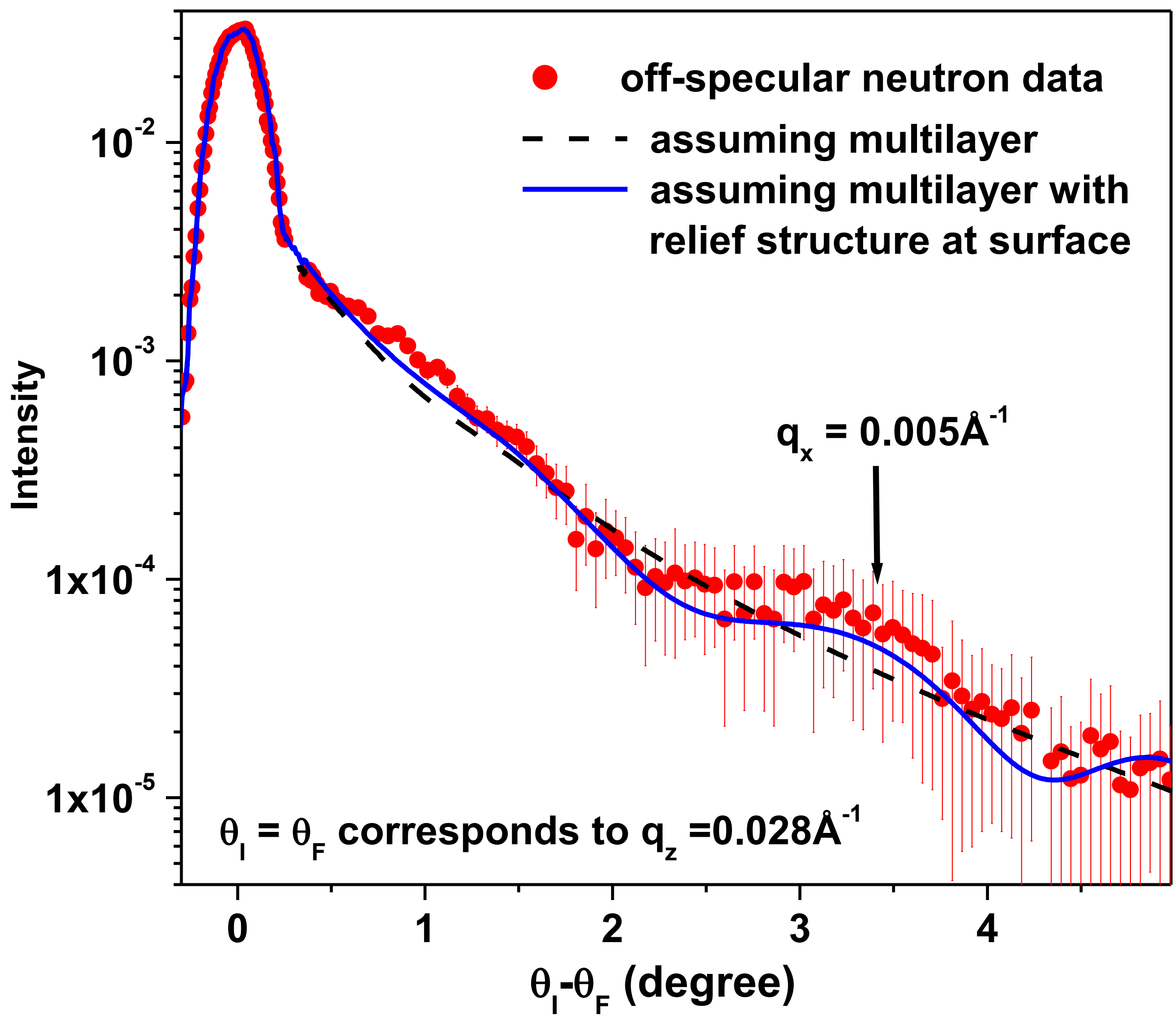
off-specular neutron data
assuming multilayer
assuming multilayer with
relief structure at surface
q_x = 0.005Å^-1
θ_I = θ_F corresponds to q_z =0.028Å^-1
Intensity
10^-2
10^-3
1x10^-4
1x10^-5
0
1
2
3
4
θ_I-θ_F (degree)

Fig 5

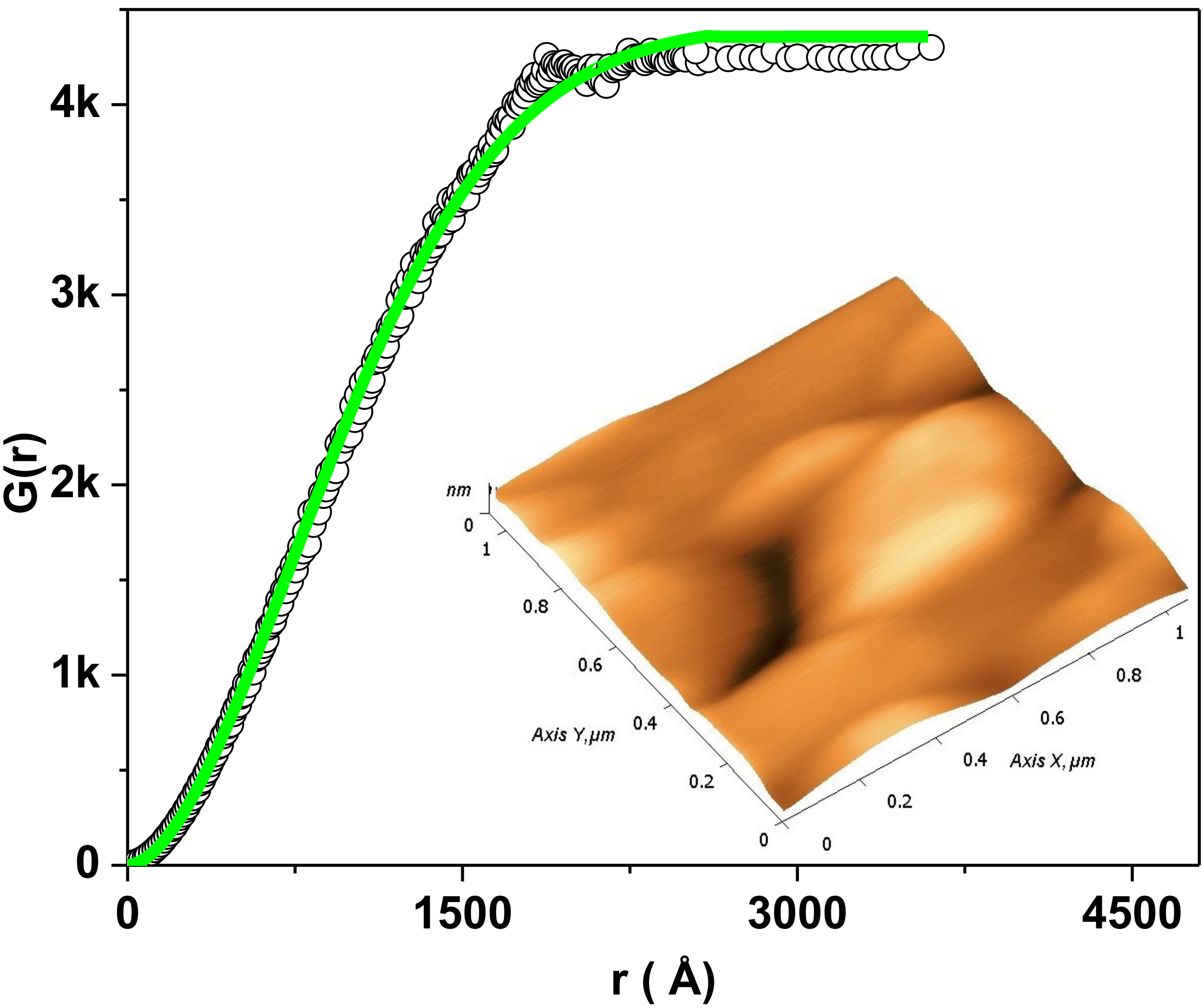

Fig. 6